\documentclass[aps,prd,superscriptaddress,
showpacs,preprintnumbers]{revtex4}
\usepackage{graphicx}

\newcommand{\be}{\begin{equation}}
\newcommand{\ee}{\end{equation}}
\newcommand{\bea}{\begin{eqnarray}}
\newcommand{\eea}{\end{eqnarray}}

\newcommand{\bfk}{\mbox{\boldmath $k$}}

\newcommand{\pup}{p^\uparrow}
\newcommand{\pdown}{p^\downarrow}

\newcommand{\bfp}{\mbox{\boldmath $p$}}
\newcommand{\bfP}{\mbox{\boldmath $P$}}

\def\lsim{\mathrel{\rlap{\lower4pt\hbox{\hskip1pt$\sim$}}\raise1pt\hbox{$<$}}}
\def\gsim{\mathrel{\rlap{\lower4pt\hbox{\hskip1pt$\sim$}}\raise1pt\hbox{$>$}}}
\def\nostrocostruttino#1\over#2{\mathrel{\mathop{\kern 0pt \rlap
{\hbox{$#1$}}} \hbox{\kern-.135em $#2$}}}

\newcommand{\NP}[1]{{\it Nucl.\ Phys.}\ {\bf #1}}

\newcommand{\PL}[1]{{\it Phys.\ Lett.}\ {\bf #1}}
\newcommand{\PR}[1]{{\it Phys.\ Rev.}\ {\bf #1}}

\begin{document}
\preprint{IC/HEP/04-4}
\title{Accessing Sivers gluon distribution via transverse single spin 
asymmetries in {\mbox{\boldmath $\pup \!\! p \to D \, X$}} processes at RHIC }
\author{M.~Anselmino}
\affiliation{Dipartimento di Fisica Teorica, Universit\`a di Torino and 
          INFN, Sezione di Torino, Via P. Giuria 1, I-10125 Torino, Italy} 
\author{M.~Boglione}
\affiliation{Dipartimento di Fisica Teorica, Universit\`a di Torino and 
          INFN, Sezione di Torino, Via P. Giuria 1, I-10125 Torino, Italy} 
\author{U.~D'Alesio}
\affiliation{INFN, Sezione di Cagliari and Dipartimento di Fisica,  
Universit\`a di Cagliari, C.P. 170, I-09042 Monserrato (CA), Italy }
\author{E.~Leader}
\affiliation{Imperial College, Prince Consort Road, London, SW7 2BW, 
England}
\author{F.~Murgia}
\affiliation{INFN, Sezione di Cagliari and Dipartimento di Fisica,  
Universit\`a di Cagliari, C.P. 170, I-09042 Monserrato (CA), Italy }
%
%\date{\today}

\begin{abstract}
\noindent
The production of $D$ mesons in the scattering of transversely polarized 
protons off unpolarized protons at RHIC offers a clear opportunity to 
gain information on the Sivers gluon distribution function. $D$ production 
at intermediate rapidity values is dominated by the elementary 
$gg \to c \bar c$ channel; contributions from $q \bar q \to c\bar c$ 
$s$-channel become important only at very large values of $x_F$. In both 
processes there is no single spin transfer, so that the final $c$ or $\bar c$ 
quarks are not polarized. Therefore, any transverse single spin asymmetry 
observed for $D$'s produced in $\pup p$ interactions cannot originate from 
the Collins fragmentation mechanism, but only from the Sivers effect in the 
distribution functions. In particular, any sizeable spin asymmetry measured 
in $\pup p \to D \, X$ at mid-rapidity values will be a direct 
indication of a non zero Sivers gluon distribution function. 
We study the $\pup p \to D \, X$ process including intrinsic transverse 
motion in the parton distribution and fragmentation functions and in the 
elementary dynamics, and show how results from RHIC could allow a measurement 
of $\Delta ^N f_{g/\pup}$. 
\end{abstract}
\maketitle
\section{Introduction and formalism}
Within the QCD factorization scheme, the cross section for an inclusive large 
$p_T$ scattering process between hadrons, like $p \, p \to h + X$, is 
calculated by convoluting the elementary partonic cross sections with the 
parton distribution functions (pdf's) and fragmentation functions (ff's). 
These objects account for the soft non-perturbative part of the scattering 
process, by giving the probability density of finding partons inside the 
hadrons (or hadrons inside fragmenting partons) carrying a specific fraction 
$x$ (or $z$) of the parent light-cone momentum. The parton intrinsic motion 
-- demanded by uncertainty principle and gluon emission -- is usually
integrated out in the high energy factorization scheme, and only partonic 
collinear configurations are considered. However, it is well known that the 
quark and gluon intrinsic transverse momenta $\bfk_\perp$ have to be taken 
into account to improve agreement with data on unpolarized cross sections at 
intermediate energies \cite{fff, pkt}. Moreover, without intrinsic 
$\bfk_\perp$ one would never be able to explain single spin asymmetries 
within QCD factorization scheme; several large single spin asymmetries 
have been observed \cite{ags,e704,star}, which in a collinear configuration 
are predicted to be either zero or negligibly small. Although not rigorously 
proven in general \cite{col,ji}, the usual factorized structure of the 
collinear scheme has been generalized with inclusion of intrinsic 
$\bfk_\perp$, so that the cross section for a generic process 
$A \,B \to C \, X$ reads:
\be
d\sigma = \sum_{a,b,c} \hat f_{a/A}(x_a,\bfk_{\perp a}) \otimes 
\hat f_{b/B}(x_b, \bfk_{\perp b}) \otimes
d\hat\sigma^{ab \to c \dots}(x_a, x_b, \bfk_{\perp a}, \bfk_{\perp b}) 
\otimes \hat D_{C/c}(z, \bfk_{\perp C}) \>.
\label{ltgen}
\ee
The pdf's and the ff's are phenomenological quantities which have to be 
obtained -- at least at some scale -- from experimental observation and 
cannot be theoretically predicted. The pdf's of unpolarized nucleons, 
$q(x) \equiv f_{q/p}(x) \equiv f_1^q(x)$, are now remarkably well known; 
one measures them in inclusive deep inelastic scattering processes at 
some scale, and, thanks to their universality and known QCD evolution, 
can use them in different processes and at different energies. The $k_\perp$
dependence of $\hat q(x, k_\perp)$ is usually assumed to be of a gaussian 
form, and the average $k_\perp$ value can be fixed so that it agrees with 
experimental data. Notice that, in our notations, a hat over a pdf or a ff 
signals its dependence on $\bfk_\perp$ and $k_\perp = |\bfk_\perp|$.

When considering polarized nucleons the number of pdf's involved grows and 
dedicated polarized experiments have to be performed in order to isolate and 
measure these functions. We have by now good data on the pdf's of 
longitudinally polarized protons -- the helicity distribution 
$\Delta q(x) \equiv g_1^q(x)$ -- but nothing is experimentally known on the 
transverse spin distribution -- the transversity function 
$\Delta_Tq(x) \equiv \delta q(x) \equiv h_1^q(x)$. 
The situation gets much more intricate when parton intrinsic transverse 
momenta are taken into account. Many more distribution and fragmentation 
functions arise, like the Sivers function 
$\Delta^N f (x, \bfk _\perp) \propto f_{1T}^\perp (x, k_\perp)$ 
\cite{siv,dp}, which describes the probability 
density of finding unpolarized partons inside a transversely polarized 
proton; similarly, the Collins fragmentation function \cite{col} gives the 
number density of unpolarized hadrons emerging in the fragmentation of a 
transversely polarized quark. These are the functions which could explain 
single spin asymmetries in terms of parton dynamics \cite{noi1,bm}.  
    
One of the difficulties in gathering experimental information on these new
spin and $\bfk_\perp$ dependent pdf's and ff's is that most often two or more 
of them  contribute to the same physical observable, making it impossible to 
estimate each single one separately.

In Ref. \cite{DY} it was shown how properly defined single spin asymmetries 
in Drell-Yan processes depend only on the Sivers distribution function
$\Delta^N f (x, \bfk_\perp)$ of quarks (apart from the usual known 
unpolarized quark distributions). In Ref. \cite{BV} it has been suggested to
look at back-to-back correlations in azimuthal angles of jets produced 
in $\pup p$ RHIC interactions in order to access the gluon Sivers 
function. We consider here another case which, again, would isolate 
the gluon Sivers effect, making it possible to reach direct independent 
information on $\Delta^N f_{g/\pup} (x, \bfk_\perp)$.

Let us consider the usual single spin asymmetry
\be
A_N = \frac{d\sigma ^\uparrow \, - \, d\sigma ^\downarrow}
           {d\sigma ^\uparrow \, + \, d\sigma ^\downarrow} \label{an}
\ee
for $\pup p \to DX$ processes at RHIC energy, $\sqrt{s} = 200$ GeV. These 
$D$ mesons originate from $c$ or $\bar c$ quarks, which at LO can be created 
either via a $q \bar q$ annihilation, $q \bar q \to c \bar c$, or via a gluon 
fusion process, $gg \to c \bar c$. The elementary cross section for 
the fusion process includes contributions from $s$, $t$ and $u$-channels,
and turns out to be much larger than the $q \bar q \to c \bar c$ 
cross section, which receives contribution from the $s$-channel alone. 
Therefore, the gluon fusion dominates the whole $\pup p \to DX$ process 
up to $x_F \simeq 0.6$. Beyond this the $q \bar q \to c \bar c$ contribution 
to the total cross section becomes slightly larger than the 
$g g \to c \bar c$ contribution, due to the much smaller values, at large
$x$, of the gluon pdf, as compared to the quark ones (see Fig.~1).

As the gluons cannot carry any transverse spin the elementary process 
$gg \to c \bar c$ results in unpolarized final quarks. In the 
$q \bar q \to c \bar c$  process one of the initial partons (that inside 
the transversely polarized proton) can be polarized; however, there is no 
single spin transfer in this $s$-channel interaction so that the final 
$c$ and $\bar c$ are again not polarized. One might invoke the possibility
that also the quark inside the unpolarized proton is polarized \cite{dan},
so that both initial $q$ and $\bar q$ are polarized: even in this case 
the $s$-channel annihilation does not create a polarized final $c$ or 
$\bar c$. Consequently, the charmed quarks fragmenting into the observed $D$ 
mesons {\it cannot be polarized}, and there cannot be any Collins 
fragmentation effect [see further comments after Eq. (\ref{final-unp})].      

Therefore, transverse single spin asymmetries in $\pup p \to DX$ can only be 
generated by the Sivers mechanism, namely a spin-$\bfk_\perp$ asymmetry in 
the distribution of the unpolarized quarks and gluons inside the polarized 
proton, coupled respectively to the unpolarized interaction process 
$q \bar q \to c \bar c$ and $gg \to c \bar c$, and
the unpolarized fragmentation function of either the $c$ or the $\bar c$ 
quark into the final observed $D$ meson. That is \cite{col,unp}:
%\goodbreak
%
\bea
d\sigma ^\uparrow - d\sigma ^\downarrow &=& 
\frac{E_D \, d\sigma^{\pup p \to DX}} {d^{3} \bfp_D} -
\frac{E_D \, d\sigma^{\pdown p \to DX}} {d^{3} \bfp_D}  
\label{final-ssa} \\
&& \hspace*{-3.0cm} = \>
\int dx_a \, dx_b  \, dz \, d^2 \bfk_{\perp a} \, d^2 \bfk_{\perp b} \, 
d^3 \bfk_{D} \, 
\delta (\bfk_{D} \cdot \hat{\bfp}_c) \, 
\delta (\hat s +\hat t +\hat u - 2m_Q^2) \> 
{\mathcal C}(x_a,x_b,z,\bfk_D) \nonumber \\
&& \hspace*{-3.0cm} \times \, \Biggl\{ \sum_q
\left[\Delta ^N f_{q/\pup}(x_a,\bfk_{\perp a}) \> \hat f_{\bar q/p}(x_b, 
\bfk_{\perp b}) \>
\frac{d \hat{\sigma}^{q \bar q \to Q \bar Q}}
{d\hat t}(x_a, x_b, \bfk_{\perp a}, \bfk_{\perp b}, \bfk_D) \>
\hat D_{D/Q}(z,\bfk_D) \right]  \nonumber \\
&& \hspace*{-3.0cm} +
\left[ \Delta ^N f_{g/\pup}(x_a,\bfk_{\perp a}) \> \hat f_{g/p}(x_b, 
\bfk_{\perp b}) \>
\frac{d \hat{\sigma}^{gg \to Q \bar Q}}
{d\hat t}(x_a, x_b, \bfk_{\perp a}, \bfk_{\perp b}, \bfk_D) \>
\hat D_{D/Q}(z,\bfk_D) \right] \Biggr\} \>, \nonumber 
\eea
where $q = u, \bar u, d, \bar d, s, \bar s$ and $Q = c$ or $\bar c$, according
to whether $D= D^+, \> D^0$ or $D= D^-, \> \overline D^{\,0}$. Notice that $z$ 
is the light-cone momentum fraction along the fragmenting parton direction,
identified by $\hat{\bfp}_c$, $z=p_D^+/p_c^+$.
Throughout the paper we choose $XZ$ as the $D$ production 
plane, with the polarized proton moving along the positive $Z$-axis and the 
proton polarization $\uparrow$ along the positive $Y$-axis. In such a frame
$\bfk_{\perp a}$ and $\bfk_{\perp b}$ have only $X$ and $Y$ components, while
$\bfk_D$ has all three components; the function 
$\delta (\bfk_D \cdot \hat{\bfp}_c)$ ensures that the integral 
over $\bfk_D$ is only performed along the appropriate transverse direction,
$\bfk_{\perp D}$, that is the transverse momentum of the produced $D$ with 
respect to the fragmenting quark direction.
The factor $\mathcal C$ contains the flux and relevant Jacobian factors for 
the usual transformation from partonic to observed meson phase space, which, 
accounting for the transverse motion, reads \cite{unp,cahn}:
\be
{\mathcal C} = \frac{\hat s}{\pi z^2}\,\frac{\hat s}{x_a x_b s}\,
\frac{ \left( E_D+\sqrt{\bfp_D^2 - \bfk_{\perp D}^2} \right)^2}
{4(\bfp_D^2 - \bfk_{\perp D}^2)} \,
\left[1- \frac{z^2 m_Q^2}
{ \left( E_D+\sqrt{\bfp_D^2 - \bfk_{\perp D}^2} \right)^2}\right]^2 
\,.
\ee
Notice that for collinear and massless particles this factor reduces to  
the familiar $\hat s/\pi z^2$. 
The Sivers distribution functions \cite{siv} 
for quarks and gluons are defined by 
\be
\Delta ^N f_{a/\pup}(x_a,\bfk_{\perp a}) = 
\hat f_{a/\pup} (x_a,\bfk_{\perp a}) - \hat f_{a/\pdown}(x_a,\bfk_{\perp a}) = 
\hat f_{a/\pup} (x_a,\bfk_{\perp a}) - \hat f_{a/\pup}(x_a,-\bfk_{\perp a})\,,
\label{siv}
\ee
where $a$ can either be a light quark or a gluon. Similarly, 
$\hat D_{D/Q}(z,\bfk_{\perp D})$ is the probability density for a quark $Q$ 
to fragment into a $D$ meson with light-cone momentum fraction $z$ 
and intrinsic transverse momentum $\bfk_{\perp D}$.

The heavy quark mass $m_Q$ is taken into account in the amplitudes of both 
the partonic processes and the resulting elementary cross sections are:
\bea
\frac{d\hat\sigma ^{q\bar q\to Q\bar Q}}{d\hat t} &=& 
\frac {\pi \alpha_s^2}{\hat s^2} \, \frac {2}{9}
\left(2\tau_1^2 + 2\tau_2^2 + \rho \right) \>,
\label{qqunp} \\
\frac{d\hat\sigma ^{gg\to Q\bar Q}}{d\hat t} &=& 
\frac {\pi \alpha_s^2}{\hat s^2} \, \frac {1}{8}\,
\left(\frac{4}{3\tau_1\tau_2}-3\right)
\left(\tau_1^2 + \tau_2^2 + \rho -\frac{\rho^2}{4\tau_1\tau_2} \right) \>,
\label{ggunp} 
\eea
where $\tau _{1,2}$ and $\rho$ 
are dimensionless quantities defined in terms of the partonic Mandelstam 
variables $\hat s$, $\hat t$ and $\hat u$ as: 
\bea
\tau_1 &=& \frac{m_Q^2-\hat t}{\hat s} \,,\nonumber \\
\tau_2 &=& \frac{m_Q^2-\hat u}{\hat s} \,,          \\
\rho &=& \frac{4m_Q^2}{\hat s} \,\cdot \nonumber
\label{tau}
\eea

The denominator of $A_N$, Eq. (\ref{an}), is analogously given by 
\bea
d\sigma ^\uparrow + d\sigma ^\downarrow &=& 
\frac{E_D \, d\sigma^{\pup p \to DX}} {d^{3} \bfp_D} +
\frac{E_D \, d\sigma^{\pdown p \to DX}} {d^{3} \bfp_D}
= 2 \, \frac{E_D \, d\sigma^{pp \to DX}} {d^{3} \bfp_D} 
\label{final-unp} \\ 
&& \hspace*{-2.5cm} = 2
\int dx_a \, dx_b \, dz\, d^2 \bfk_{\perp a} \, d^2 \bfk_{\perp b} \, 
d^3 \bfk_D \, 
\delta (\bfk_D \cdot \hat{\bfp}_c) \, 
\delta (\hat s +\hat t +\hat u - 2m_Q^2) \, {\mathcal C}(x_a,x_b,z,\bfk_D)
\nonumber \\
&&  \hspace*{-2.5cm} \times \, \Biggl\{ \sum_q
\left[ \hat f_{q/p}(x_a,\bfk_{\perp a}) \> 
\hat f_{\bar q/p}(x_b, \bfk_{\perp b}) \>
\frac{d \hat{\sigma}^{q \bar q \to Q \bar Q}}
{d\hat t}(x_a, x_b, \bfk_{\perp a}, \bfk_{\perp b},  \bfk_D) \>
\hat D_{D/Q}(z,\bfk_D) \right] \nonumber \\
&& \hspace*{-2.5cm} +
\left[ \hat f_{g/p}(x_a,\bfk_{\perp a}) \> \hat f_{g/p}(x_b, \bfk_{\perp b}) \>
\frac{d \hat{\sigma}^{gg \to Q \bar Q}}
{d\hat t}(x_a, x_b, \bfk_{\perp a}, \bfk_{\perp b}, \bfk_D) \>
\hat D_{D/Q}(z,\bfk_D) \right] \Biggr\} \>. \nonumber 
\eea

In Eqs. (\ref{final-ssa}) and (\ref{final-unp}) we consider intrinsic 
transverse motions in the distributions of initial light quarks, in the 
elementary process and in the heavy quark fragmentation function, {\it i.e.} 
we consider a fully non planar configuration for the partonic scattering.  
This has two main consequences: on one side, taking into account three 
intrinsic transverse momenta makes both the kinematics and the dynamics highly 
non trivial; on the other side it generates a large number of contributions, 
other than the Sivers effect, which originate from all possible combinations 
of $\bfk_\perp$ dependent distribution and fragmentation functions, each 
weighted by a phase factor (given by some combination of sines and cosines of 
the azimuthal angles of the parton and final meson momenta). This 
topic deserves a full treatment on its own, which will soon be presented in 
Refs.~\cite{prep1,prep2}. At present, we only point out that we have 
explicitely verified that all contributions to the $\pup p \to DX$ single 
spin asymmetry from $\bfk_\perp$ dependent pdf's and ff's, aside from those 
of the Sivers functions, are multiplied by phase factors which make the 
integrals over the transverse momenta either negligibly small or 
identically zero. Therefore, they can be safely neglected here.

Eq. (\ref{final-ssa}) shows how $A_N$ depends on the unknown Sivers 
distribution function; as all other functions contributing to $A_N$ are 
reasonably well known (including the fragmentation function $D_{D/Q}$ 
\cite{cacc}) a measurement of $A_N$ should bring direct information on 
$\Delta ^N f_{g/\pup}$ and, to some extent, $\Delta ^N f_{q/\pup}$.

\section{Numerical estimates}

\begin{center} 
\begin{figure}[t]
\includegraphics[width=7.5cm,angle=-90]{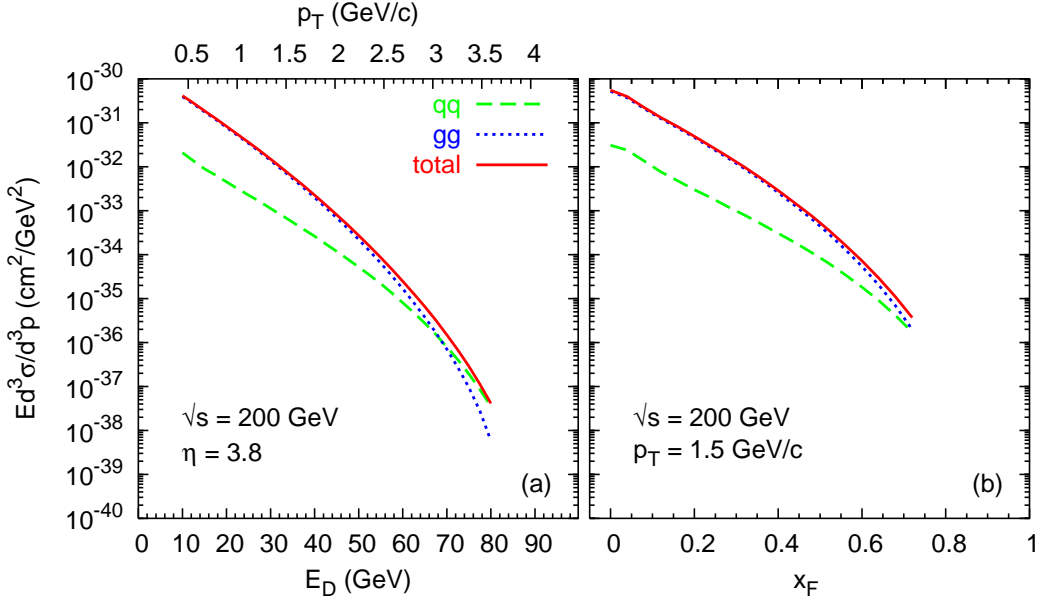}
\caption{\label{unp-sigma} \small{
The unpolarized cross section for the process $pp \to DX$ at $\sqrt s = 200$ 
GeV, as a function of $E_D$ and $p_T$ at fixed pseudo-rapidity $\eta=3.8$ (a),
and as a function of $x_F$ at fixed transverse momentum $p_T=1.5$ GeV/$c$ (b), 
calculated according to Eqs.~(\ref{final-unp}) and (\ref{unp-pdf}). 
The solid line is the full cross section, whereas the dashed and dotted lines 
show the $q\bar q \to c\bar c$ and $gg \to c\bar c$ contributions separately.
}}
\end{figure}
\end{center}

So far, all analyses and fits of the single spin asymmetry data were 
based on the assumption that the gluon Sivers function $\Delta ^N f_{g/\pup}$ 
is zero. RHIC data on $A_N$ in $\pup p \to DX$ will enable us to test the 
validity of this assumption.
In fact, as the $gg\to c\bar c$ elementary scattering largely dominates the 
process up to $x_F\simeq 0.6$ (see Fig.~1), any sizeable single spin asymmetry 
measured in $\pup p \to DX$ at moderate $x_F$'s would be the direct 
consequence of a non zero contribution of $\Delta ^N f_{g/\pup}$. 
For $x_F \gsim 0.6$ the competing $q \bar q\to c\bar c$ term becomes 
approximately the same size as $gg\to c \bar c$ (see Fig.~\ref{unp-sigma}); 
consequently the quark and gluon Sivers functions could contribute to $A_N$ 
in approximately equal measure making the data analysis more involved, as we 
shall discuss below.

Since we have no information about the gluon Sivers function from other 
experiments, we are unable to give predictions for the size of the $A_N$ 
one can expect to measure at RHIC. Instead, we show what asymmetry one can 
find in two opposite extreme scenarios: the first being the case in which 
the gluon Sivers function is set to zero, 
$\Delta ^N f_{g/\pup}(x_a,\bfk_{\perp a}) = 0$, and the quark Sivers function 
$\Delta ^N f_{q/\pup}(x_a,\bfk_{\perp a})$ is taken to be at its maximum 
allowed value at any $x_a$; the second given by the opposite situation, where 
$\Delta ^N f_{q/\pup}=0$ and $\Delta ^N f_{g/\pup}$ is maximized in $x_a$.

Concerning the $\bfk_{\perp}$ dependence of the unpolarized pdf's and 
the Sivers functions, we adopt, both for quarks and gluons, a most natural 
and simple factorized Gaussian parameterization 
\bea
&&\hat f(x,\bfk_\perp) = f(x) \, \frac{1}{\pi \langle k_\perp^2 \rangle} \, 
e^{-k_\perp ^2/\langle k_\perp^2 \rangle} \>, \label{unp-pdf} \\
&&\Delta ^N f(x,\bfk_\perp) =  \Delta ^N f(x) \,
\frac{1}{\pi \langle k_\perp^2 \rangle} \, 
e^{-k_\perp ^2/\langle k_\perp^2 \rangle } \, 
\frac{2 k_\perp M}{k_\perp^2 + M^2} \, \cos(\phi_{k_\perp})\,, 
\label{siv-pdf} 
\eea
where $M = \sqrt{\langle k_\perp^2 \rangle}$ and 
$\phi_{k_\perp}$ is the $\bfk _\perp$ azimuthal angle. The extra factor 
$2 k_\perp M/(k_\perp^2 + M^2)$ in the Sivers function is chosen in such 
a way that, while ensuring the correct small $k_\perp$ behaviour, it equals 
1 at $k_\perp = M$, being always smaller at other values. The azimuthal  
$\cos(\phi_{k_\perp})$ dependence is the only one allowed by Lorentz 
invariance, via the mixed vector product $\bfP \cdot (\hat{\bfp} \times
\hat{\bfk}_\perp)$ where, with our frame choice, $\bfP = (0,1,0)$ is
the proton polarization vector, $\hat{\bfp} = (0,0,1)$ is the unit vector along
the polarized proton motion and $\hat{\bfk}_\perp = (\cos(\phi_{k_\perp}), 
\sin(\phi_{k_\perp}), 0)$. 

The Sivers functions (\ref{siv-pdf}) for both quarks and gluons must respect 
the positivity bound
\be
\frac{|\Delta^N f_{a/\pup}(x_a, \bfk_{\perp a})|}
{2 \,\hat f_{a/p}(x_a, k_{\perp a})}  \leq 1
\quad\quad \forall\, x_a, \; k_{\perp a} \,,
\label{posb}
\ee
which means that Eq. (\ref{posb}) can be satisfied for any $x_a$ and 
$\bfk_{\perp a}$ values by taking 
\be
\Delta^N f_{a/\pup}(x_a) \leq 2\,f_{a/p}(x_a)\,. \label{sat}
\ee
%

%%%%%%%%%%%%%%%%%%%%%%%%%%%%%%%%%%%%%%%%%%%%%%%%%%%%%%%%%%%%%%%%%%%%%%%%%%%
%%%%%%%%%%%%%%%%%%%%%%%%%%%%%%%%%%%%%%%%%%%%%%%%%%%%%%%%%%%%%%%%%%%%%%%%%%%

For the fragmentation function $\hat D_{D/Q}(z,\bfk_{\perp D})$ we adopt a 
similar model, in which we assume factorization of $z$ and $\bfk_{\perp D}$ 
dependences 
\be
\hat D_{D/Q}(z,\bfk_{\perp D}) = D_{D/Q}(z)\,g(\bfk_{\perp D})\,,
\ee
where $D_{D/Q}(z)$ is the usual fragmentation function available in the 
literature (see for instance Ref.~\cite{cacc}) and $g(\bfk_{\perp D})$ is a 
gaussian function of $|\bfk_{\perp D}|^2$ analogous to that in 
Eq.~(\ref{unp-pdf}), normalized so that, for a fragmenting quark of momentum 
$\bfp_c$, 
\be
\int d^3\bfk_D \, \delta(\bfk_D \cdot \hat{\bfp}_c) 
\, \hat D_{D/Q}(z,\bfk_D) = D_{D/Q}(z)\,.
\ee
%
%%%%%%%%%%%%%%%%%%%%%%%%%%%%%%%%%%%%%%%%%%%%%%%%%%%%%%%%%%%%%%%%%%%%%%%%%%%
\begin{center} 
\begin{figure}[t]
\includegraphics[width=7.5cm,angle=-90]{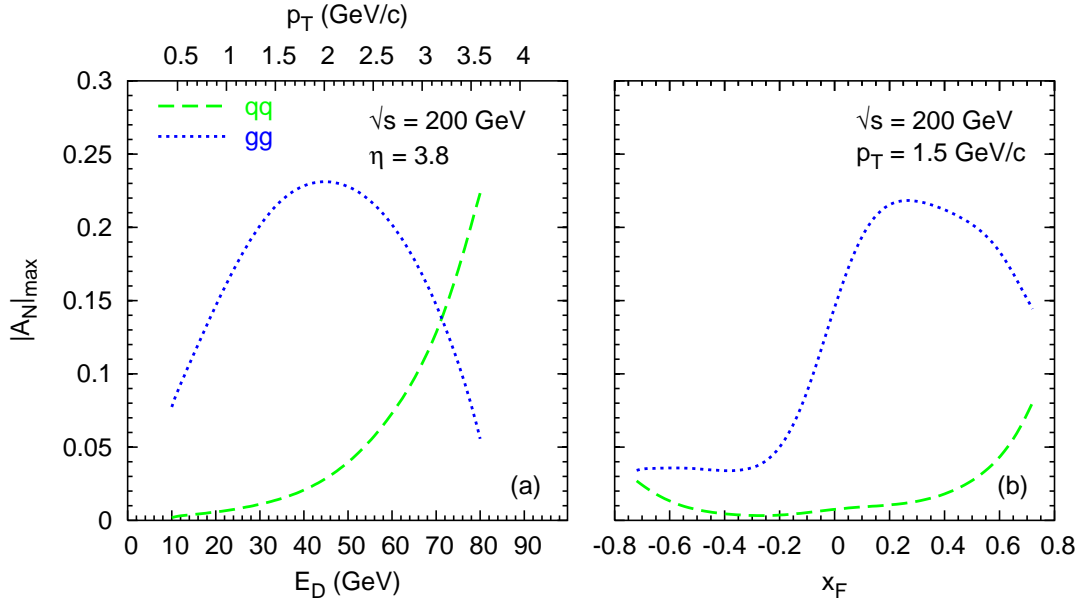}
\caption{\label{A-N} \small{
Maximized values of $|A_N|$ for the process $\pup p \to DX$ as a function of 
$E_D$ and $p_T$ at fixed pseudo-rapidity (a), and as a function of $x_F$ at 
fixed transverse momentum (b), calculated using saturated Sivers functions, 
according to Eq. (\ref{sat}) of the text. The dashed line corresponds to a 
maximized quark Sivers function (with the gluon Sivers function set to zero), 
while the dotted line corresponds to a maximized gluon Sivers function (with 
the quark Sivers function set to zero).
}}
\end{figure}
\end{center}
%%%%%%%%%%%%%%%%%%%%%%%%%%%%%%%%%%%%%%%%%%%%%%%%%%%%%%%%%%%%%%%%%%%%%%%%%%%

\vspace*{-13pt}

In Fig.~\ref{unp-sigma}(a) we show the unpolarized cross section for the 
process $p\,p \to DX$ at $\sqrt s = 200$ GeV as a function of both the 
heavy meson energy $E_D$ and its transverse momentum $p_T$, at fixed 
pseudo-rapidity $\eta = 3.8$ (notice that $x_F \simeq E_D/$(100 GeV)). 
In Fig.~\ref{unp-sigma}(b) the same total cross section is presented as 
a function of $x_F$ at fixed $p_T = 1.5$ GeV/$c$. The $x$ and 
$Q^2$-dependent parton distribution functions $f _{q/p} (x, Q^2)$ are taken 
from MRST01~\cite{mrst01}, while the $k_\perp$ dependence is fixed by  
Eq. (\ref{unp-pdf}) with $\sqrt{\langle k_\perp^2 \rangle} = 
0.8$ GeV/$c$ \cite{unp}; similarly, the fragmentation functions 
$D_{D/Q} (z, Q^2)$ are from Ref.~\cite{cacc}, with the $k_\perp$ dependence
fixed by $\sqrt{\langle k_\perp^2 \rangle} = 0.8$ GeV/$c$. We have 
explicitely checked that our numerical results have very 
little dependence on the $\langle k_\perp^2 \rangle$ value of the 
fragmentation functions. Finally, we have taken as QCD scale $Q^2 = m_Q^2$.   
The dashed and dotted lines correspond to the $q\bar q \to c\bar c$ and 
$gg \to c\bar c$ contributions respectively, whereas the solid line gives 
the full unpolarized cross section. 
These plots clearly show the striking dominance of the $gg \to c\bar c$ 
channel over most of the $E_D$ and $x_F$ ranges covered by RHIC kinematics. 

Fig.~\ref{A-N} shows our estimates for the maximum value of the single spin 
asymmetry in $\pup p \to DX$. The dashed line shows $|A_N|$ when the quark 
Sivers function is set to its maximum, {\it i.e.} 
$\Delta ^N f _{q/\pup}(x) = 2f _{q/p}(x)$, 
while setting the gluon Sivers function to zero. 
Clearly, the quark contribution to $A_N$ is very small over most of the 
kinematic region, at both fixed pseudo-rapidity and varying $E_D$, Fig. 2(a), 
and fixed $p_T$ and varying $x_F$, Fig. 2(b).
The dotted line corresponds to the SSA one finds in the opposite situation, 
when $\Delta ^N f _{g/\pup}(x) = 2f _{g/p}(x)$ and $\Delta ^N f _{q/\pup} =0$:
in this case the asymmetry presents a sizeable maximum in the central $E_D$ 
and positive $x_F$ energy region (in our configuration positive $x_F$ means 
$D$ mesons produced along the polarized proton direction, {\it i.e.} the 
positive $Z$-axis). This particular shape is given by the azimuthal 
dependence of the numerator of $A_N$, see Eqs.~(\ref{final-ssa}) and 
(\ref{siv-pdf}). When the energy $E_D$ is small, $p_T$ is also very small 
(for instance, for $E_D \leq 23$ GeV, $p_T \leq 1$ GeV/$c$) and the partonic 
cross sections $d\hat \sigma/d\hat t$ depend only very weakly on 
$\phi_{k_{\perp a}}$. Therefore, when we integrate over $\phi_{k_{\perp a}}$ 
the partonic cross sections multiplied by the factor 
$\cos(\phi_{k_{\perp a}})$ from the Sivers function, we obtain 
negligible values. The transverse momentum $p_T$ of the detected $D$ meson 
grows with increasing $E_D$ and the partonic cross sections become more and 
more sensitively dependent on $\phi_{k_{\perp a}}$: then $A_N$ grows and a 
peak develops in correspondence of $\phi_{k_{\perp a}} \simeq 0$. Similarly,
one can understand the behaviour of $|A_N|_{\rm max}$ in the negative
and positive $x_F$ regions, Fig. 2(b). Only at 
very large $E_D$ and $x_F$ the $q\bar q \to c\bar c$ contribution becomes 
important, and a rigorous analysis in that region
will only be possible when data from independent sources will provide enough 
information to be able to separate the two contributions.  

By looking at Fig.~\ref{A-N} it is natural to conclude that any sizeable 
single transverse spin asymmetry measured by STAR or PHENIX experiments
at RHIC in the region $E_D\le 60$ GeV or $-0.2 \le x_F \le 0.6$, would give 
direct information on the size and importance of the gluon Sivers function.

\section{Comments and conclusions}

We have shown that the observation of the transverse single spin asymmetry 
$A_N$ for $D$ mesons generated in $\pup p$ scattering 
offers a great chance to study the Sivers distribution functions. 
This channel allows a direct, uncontaminated access to this function since 
the underlying elementary processes guarantee the absence of any 
polarization in the final partonic state; consequently, 
contributions from Collins-like terms cannot be present to influence the 
measurement. Moreover, the large dominance of the $gg \to c \bar c$ process 
at low and intermediate $x_F$ offers a unique opportunity to measure the 
gluon Sivers distribution function $\Delta ^N f _{g/\pup}$.

Once more, intrinsic parton motions play a crucial role and have to be 
properly taken into account. Adopting a simple model to parameterize the
$k_\perp$ dependence we have given some estimates of the unpolarized 
cross section for $D$ meson production, together with some upper 
estimate of the SSA in the two opposite scenarios in which either 
$\Delta ^N f _{g/\pup}$ is maximal and $\Delta ^N f _{q/\pup} = 0$ or 
$\Delta ^N f _{g/\pup}=0$ and $\Delta ^N f _{q/\pup}$ is maximal. 
Our results hold for $D = D^+, D^-, D^0, \overline D^{\,0}$.   
Both the cross section and $A_N$ could soon be measured at RHIC.

It clearly turns out that any sizeable contributions to the 
$\pup p \to DX$ single spin asymmetry at low to intermediate $E_D$'s or 
$x_F$'s would be a most direct indication of a non zero gluon Sivers 
function.

\begin{acknowledgments}
We thank C. Aidala and L. Bland for interesting and inspiring 
discussions.
The authors are grateful to INFN for continuous support to their 
collaboration.
M. Boglione is grateful to Or\'eal Italia for awarding her a 
grant from the scheme ``For Women in Science''. E. Leader is grateful 
to the Royal Society of Edinburgh Auber Bequest for support. U.D. and
F.M. acknowledge partial support from ``Cofinanziamento MURST-PRIN03''.
\end{acknowledgments}

\end{document}